\documentstyle[prl,aps,epsf,multicol]{revtex}
\begin{document}
\draft
\preprint{}
\title{The spin quantum Hall effect in unconventional superconductors}
\author{T. Senthil$^1$, J. B. Marston$^{1,2}$, and Matthew P. A. Fisher$^1$}
\address{$^1$Institute for Theoretical Physics, University of California,
Santa Barbara, CA 93106--4030 \\
$^2$\cite{permaddr}Department of Physics, Brown University, Providence, RI 
02912-1843}

\date{\today}
\maketitle

\begin{abstract}
We study the properties of  the ``spin quantum Hall
fluid" - a novel spin phase with quantized spin Hall
conductance that is potentially realizable in superconducting systems
with unconventional pairing symmetry. A simple realization is provided
by a $d_{x^2-y^2} + id_{xy}$ superconductor which we argue has
a dimensionless spin Hall conductance equal to two. A theory of the edge states
of the $d_{x^2-y^2}+id_{xy}$ superconductor is developed. The properties of the
transition to a phase with vanishing spin Hall conductance induced by disorder
are considered. We construct a description of this transition in terms of a
supersymmetric spin chain, and use it to numerically determine universal 
properties of the transition.
We discuss various possible experimental probes of this quantum Hall physics.

\end{abstract}
\vspace{0.15cm}


\begin{multicols}{2}
\narrowtext

\section{Introduction}
A remarkable property of a singlet superconductor is
the occurrence of the phenomenon of spin-charge
separation\cite{Rokh1,BFN}. The superconducting condensate
may be viewed as a collection of spinless,
charge $2e$ Cooper pairs that have Bose condensed.
The spin on the other hand is carried entirely by the
fermionic quasiparticle excitations which do not carry
definite charge. This observation is particularly
important in the context of superconductors with
non-$s$-wave Cooper pairing leading possibly to
quasiparticle excitations at arbitrarily low energies.
The best studied case is $d_{x^2 -y^2}$ pairing in the
high-$T_c$ cuprates. The resulting superconducting state has gapless
quasiparticle excitations which dominate the low temperature
properties. The cuprates thus provide an opportunity to
explore the low energy properties of a gapless spin-charge
separated system in dimensions greater than one.
Recent work\cite{short,dos} has pointed out the possibility of realizing a
novel spin phase - the ``spin metal" - in the cuprates in the presence of
disorder. This phase is characterized by a non-vanishing finite spin diffusion
constant and spin susceptibility at zero temperature, and is not
known to exist in insulating Heisenberg spin models. In this work, we explore
another novel spin phase potentially realizable in superconducting systems
- the ``spin quantum Hall fluid". This phase is characterized by
a quantized value of the Hall spin conductance (analogous to
the quantized Hall charge conductance in the integer quantum hall effect).

We begin by showing that such a spin quantum Hall fluid phase is realized
by two dimensional superconductors with
$d_{x^2-y^2} + id_{xy}$ symmetry. The $d+id$ state, which has received a fair
amount of attention recently\cite{Rokh2,Laugh,Bltsky,Sauls},
has been known to possess various similarities
with quantum Hall states though the precise characterization in terms of
spin transport has not been pointed out before. In particular, it has been
suggested that a transition from the $d_{x^2 - y^2}$ to the
$d_{x^2 - y^2} + id_{xy}$ superconductor may be driven by external
magnetic fields\cite{Laugh}, and hence is potentially realizable in the
cuprates.

Here we first calculate the bulk spin Hall conductance of the $d+id$ state
and show explicitly that it is
quantized to be equal to two (in units of the dimensionless spin conductance).
We then use semiclassical arguments to show the existence of two
spin-current carrying edge states as required by the quantization of the bulk
Hall spin conductance. A Hamiltonian describing the propagating edge modes is
derived.  We next consider the effects of disorder on the
$d+id$ state.  The quantization of the spin Hall conductance is robust to weak
impurity scattering.
However if the impurity scattering is sufficiently strong, there can be a phase
transition to a phase with vanishing Hall spin conductance.
The properties of this transition are considered next. Ignoring the
quasiparticle interactions, this transition is argued to be described by the
critical point of a replica non-linear sigma model theory\cite{short}
with a topological term which describes quasiparticle localization in a
superconductor
without time reversal but with spin rotation invariance (class C of
Ref. \onlinecite{AZ}).  We then construct a network model\cite{CC}
describing this transition, and show that it is identical to that simulated
recently by
Kagalovsky et al.\cite{John}. We then motivate a description of this transition
in terms of a supersymmetric (SUSY) spin chain. In contrast to
the SUSY spin chain which describes the usual integer quantum Hall 
transition\cite{Zirn,Tsai}, this SUSY chain has only a finite number, three, of
degrees of freedom at each site. This enables the efficient use of a numerical
technique - the density matrix renormalization group (DMRG) - which has been 
successfully used for accurate calculations of the properties of quantum spin 
chains in other situations\cite{White}. We present numerical results for
a number of universal critical properties of the transition.  Some of these 
have been obtained before from the network model simulations\cite{John}.
Very recently, Gruzberg, Ludwig and Read\onlinecite{Ilya} 
have provided a mapping
of this transition to classical percolation and determined exact
values for various critical exponents. Our numerical results are 
in excellent agreements with these exact values.
We conclude with a general discussion of various
experimental probes of the physics discussed in this paper.

\section{Bulk spin Hall conductance of the $d+id$ superconductor}
We begin by defining the spin Hall conductance. In general, the spin conductance
measures the spin current induced in the system in response to a spatially
varying Zeeman magnetic field. The spin Hall conductance measures the spin 
current in a direction transverse to the direction of variation of the external
Zeeman field.  More precisely,
a Zeeman field $B^z(y)$ along, say the $z$-direction of spin, which depends
only on, for instance, the
spatial $y$-direction leads to a current $j^z_x$ of the $z$ component of the
spin along the spatial $x$-direction given by
\begin{equation}
j^z_x = \sigma_{xy} ^s \left(-\frac{dB^z(y)}{dy}\right)
\end{equation}
with $\sigma_{xy}^s$ being the spin Hall conductance. (Note that the analog of
the ``electric" field for spin transport is the derivative of the Zeeman field).
Just like the usual Hall effect, $\sigma_{xy}^s$ is zero in the presence of
parity and time reversal invariances.
The $d+id$ superconductor is neither parity nor time reversal invariant
and hence can have a non-vanishing $\sigma_{xy}^s$.

Before proceeding further, it is worthwhile to recall some general properties
of singlet superconductors with no time reversal invariance. Consider a
general lattice BCS Hamiltonian for such a superconductor:
\begin{equation}
\label{BCS_c}
{\cal H} = \sum_{i,j}\left[t_{ij}\sum_{\alpha} c^{\dagger}_{i\alpha} c_{j\alpha}
+ \Delta_{ij} c^{\dagger}_{i\uparrow} c^{\dagger}_{\downarrow j} +
\Delta^{*}_{ij} c_{j\downarrow} c_{i\uparrow} \right]
\end{equation}
where $i,j$ refer to the sites of some lattice.
Hermiticity implies $t_{ij} = t^{*}_{ji}$,
and spin rotation invariance requires $\Delta_{ij} = \Delta_{ji}$.

It is often useful to use an alternate representation in terms of a new set of
$d$-operators defined by:
\begin{equation}
\label{c_to_d}
d_{i\uparrow} = c_{i\uparrow},~~ d_{i\downarrow} = c^{\dagger}_{i\downarrow } .
\end{equation}
The Hamiltonian, Eq. \ref{BCS_c}, then takes the form
\begin{equation}
\label{BCS_d}
{\cal H} = \sum_{ij} d^{\dagger}_i \left(\begin{array}{cc}t_{ij} & \Delta_{ij}
\\
                                                \Delta_{ij}^{*} & -t_{ij}^{*}
                                                               \end{array}
\right) d_j
                                     \equiv \sum_{ij}d_{i}^{\dagger}H_{ij}d_j .
\end{equation}
Writing $t_{ij} = a^{z}_{ij}+ib_{ij},~~\Delta_{ij} = a^{x}_{ij}-ia^{y}_{ij}$
with $\vec{a}_{ij} = \vec{a}_{ji}$, real symmetric and $b_{ij} = -b_{ji}$, real
antisymmetric gives
\begin{equation}
H_{ij} = ib_{ij} + \vec{a}_{ij} \cdot \vec{\sigma} ,
\end{equation}
where $\vec{\sigma}_i$ are the three Pauli matrices.
Note that $SU(2)$ spin rotational invariance requires
\begin{equation}
\label{d_su2}
\sigma_y H_{ij}\sigma_y = -H^{*}_{ij}
\end{equation}
Equivalently, we may require that the second quantized Hamiltonian ${\cal H}$
in Eq. \ref{BCS_d} be invariant under
\begin{equation}
\label{d_SU2}
d \rightarrow i \sigma_y d^{\dagger} .
\end{equation}

The advantage of going to the $d$ representation is that the Hamiltonian
conserves the number of
$d$ particles. Note that the transformation Eq. \ref{c_to_d} implies that the
number of
$d$ particles is essentially the $z$ component of the physical spin density:
\begin{equation}
S^z_i = \frac{\hbar}{2}\left(d^{\dagger}_id_i -1 \right) .
\end{equation}
A spin rotation about the $z$ axis corresponds to a $U(1)$ rotation of the $d$
operators.
This $U(1)$ is clearly present in the $d$ Hamiltonian. Invariance under
spin rotations about the $x$ or $y$ axes is not manifest though.

Now consider the particular case of a $d_{x^2-y^2}+id_{xy}$
superconductor\cite{note_1}.  In momentum space, the Hamiltonian is
\begin{equation}
\label{BCS}
{\cal H} = \sum_{k\alpha} \epsilon_k c^{\dagger}_{k \alpha} c_{k\alpha} +
\left( \Delta_k c^{\dagger}_{k \uparrow } c^{\dagger}_{-k \downarrow} + H.c. 
\right)
\end{equation}
where $\epsilon_k $ is the band dispersion and
$\Delta_k = \Delta_0 \cos(2\theta_k) -i\Delta_{xy} \sin(2\theta_k)$
with $\tan(\theta_k) = k_y / k_x$.
It is sometimes useful to think in terms of a lattice version of the $d+id$
superconductor.
This has been formulated by Laughlin\cite{Laugh}. Translating to momentum
space, for a square
lattice, we have $\epsilon_k \sim (\cos(k_x) + \cos(k_y)),
\Delta_k \sim \Delta_0(\cos(k_x) - \cos(k_y)) - 
i\Delta_{xy} \sin(k_x) \sin(k_y)$
which has the same symmetry under four-fold rotations of the lattice as the form
written down earlier.

The parameter $\Delta_{xy}$ measures the relative strength of the $d_{xy}$
and $d_{x^2 - y^2}$ components. $\Delta_{xy} = 0$ corresponds to the
familiar $d_{x^2 - y^2}$ state. In this limit, the gap function $\Delta_k$
vanishes at four points of the Fermi surface and there are
gapless quasiparticle excitations at these four nodes. A low energy theory of
the $d_{x^2 -y^2}$ superconductor can be obtained\cite{BFN} by linearizing the
dispersion relation of these quasiparticles around the nodes. We put
$\Upsilon_1(k) = c_k, \Upsilon_2(k) = i\sigma_y c^{\dagger}_{-k}$ for $k_y > 0$
to write
\begin{equation}
{\cal H} = {\sum_k}^\prime \Upsilon^{\dagger}(k) (\epsilon_k \tau_z +
\Delta_k \tau_x)\Upsilon(k)
\end{equation}
where the prime indicates a sum over $k_y > 0$ and $\vec{\tau}$ are the Pauli
matrices in $\Upsilon_1, \Upsilon_2$ (particle-hole) space.
If $(K_1, K_2)$ are the two nodal directions with $k_y > 0$,
we may just keep modes near $(K_1, K_2)$. Linearizing $\epsilon_k$ and
$\Delta_k$ near the nodes,
we get the following low energy theory for the $d_{x^2 - y^2}$ superconductor:
\begin{equation}
\label{Dirac1}
{\cal H} = \int d^2 X \psi^{\dagger}_1(-iv_F \partial_X \tau_z +
iv_{\Delta}\partial_Y \tau_x)\psi_1
+ (1 \leftrightarrow 2; X \leftrightarrow Y)
\end{equation}
Here $X = \frac{1}{\sqrt{2}}(x+y)$ and $Y = \frac{1}{\sqrt{2}}(-x+y)$. The
field
$\psi_i$ is the Fourier transform of $\psi_i(k) = \Upsilon(K_i + k)$ for $i =
1, 2$.
Each $\psi_i$ thus has four components $\psi_{ia\alpha}$ where $a$ is the
particle-hole index and $\alpha$ the spin index. The $\psi_i$
transform as spinors under $SU(2)$ spin rotations.
This Hamiltonian is manifestly invariant under spin $SU(2)$. (It also has
additional
$U(1)$ symmetries that can be related to momentum conservation that holds
in clean systems\cite{BFN}). The physical charge density is of course not
conserved as is already apparent from Eq. \ref{BCS}.

It is useful at this stage to express the original real space electron
operators in terms of the low energy continuum fields. This is easily seen to be
\begin{equation}
c_\uparrow(\bbox{x})
\sim e^{i\bbox{K_j \cdot \bbox{x}}} \psi_{j1\uparrow} -
e^{-i\bbox{K_j \cdot \bbox{x}}} \psi^{\dagger}_{j2 \downarrow }   ,
\end{equation}
\begin{equation}
c_{\downarrow}(x)  \sim e^{i\bbox{K_j \cdot \bbox{x}}} \psi_{j1\downarrow} +
e^{-i\bbox{K_j \cdot \bbox{x}}} \psi^{\dagger}_{j2 \uparrow} ,
\end{equation}
with a sum over the node index ($j=1,2$) understood.

Now consider introducing a small $id_{xy}$ component, {\it i.e} letting
$\Delta_{xy} > 0$. For small $\Delta_{xy}$,
we may work with the low energy theory Eq. \ref{Dirac1} near the nodes of the
$d$-SC. The
$id$ perturbation adds to the low energy Hamiltonian Eq. \ref{Dirac1} the
following term
\begin{equation}
\label{DiracM}
{\cal H}_{id} = \int d^2X \Delta_{xy} (\psi_1^{\dagger} \tau_y \psi_1 -
\psi_2^{\dagger} \tau_y \psi_2)
\end{equation}
Note that this is basically a mass term for the two Dirac theories describing
the two nodes.

The spin density can be expressed in terms of the continuum fields as
\begin{equation}
\vec{S} = \frac{\hbar}{2} \psi^{\dagger} \vec{\sigma} \psi
\end{equation}
Similarly the spin currents may also be obtained from Noether's theorem.

We now perform the continuum version of the transformation Eq. \ref{c_to_d} by
defining new fields
$\chi_{ia\alpha}$ through
\begin{eqnarray}
\psi_{ia\uparrow}    & =   & \chi_{ia\uparrow} , \\
\psi_{ia\downarrow}  & =   & \chi^{\dagger}_{ia \downarrow } .
\end{eqnarray}
The form of the Hamiltonian Eqs. (\ref{Dirac1}) and (\ref{DiracM}) is
unchanged under the transformation to the $\chi$ fields.
It is clear that the $z$-component of the physical spin density is essentially
the density of the $\chi$ particles.
A spin rotation about the $z$ axis corresponds to a $U(1)$ rotation of the 
$\chi$ fields.
This $U(1)$ is clearly present in the $\chi$ Hamiltonian. Once again invariance
under spin rotations about the $x$ or $y$ axes is not manifest.

The $d$ operator in real space may also be expressed in terms of these
continuum fields as
\begin{equation}
\label{d_to_chi1}
c_\uparrow(\bbox{x}) \equiv d_\uparrow(\bbox{x})
\sim e^{i\bbox{K_j \cdot \bbox{x}}} \chi_{j1\uparrow} -
e^{-i\bbox{K_j \cdot \bbox{x}}} \chi_{j2\downarrow}   ,
\end{equation}
\begin{equation}
\label{d_to_chi2}
c^{\dagger}_{\downarrow}(x) \equiv d_\downarrow(\bbox{x}) 
\sim e^{-i\bbox{K_j \cdot \bbox{x}}} \chi_{j1\downarrow} +
e^{i\bbox{K_j \cdot \bbox{x}}} \chi_{j2\uparrow}   ,
\end{equation}
with a sum over the node index ($j=1,2$) understood.
Note that the symmetry transformation Eq. \ref{d_SU2} implies symmetry of the
Hamiltonian under
\begin{equation}
\label{chi_SU2}
\chi_{ja\alpha} \rightarrow i(\sigma_y)_{\alpha \beta} 
\chi^{\dagger}_{ja \beta } {}.
\end{equation}

The calculation of the spin Hall conductance is simplified by choosing the
external Zeeman field to be oriented
along the $z$ spin direction. In that case, the spin Hall conductance is just
the charge Hall conductance of the $\chi$ fields. The result is
well-known\cite{LFSG}: The contribution of each Dirac species is
\begin{displaymath}
\frac{1}{2} {\rm sgn}(\Delta_{xy})\frac{(\hbar/2)^2}{2\pi\hbar}
\end{displaymath}
We have introduced the quantum of
spin conductance $\frac{(\hbar/2)^2}{2\pi\hbar} = \frac{\hbar}{8\pi}$.
As there are now four Dirac species, we obtain for the spin Hall conductance 
(in units of $\frac{\hbar}{8\pi}$) of the $d+id$ superconductor:
\begin{equation}
\label{shc}
\sigma_{xy}^s = 2~ {\rm sgn}(\Delta_{xy}) .
\end{equation}
This is the main result of this section. (If we repeat the calculation for a
$d_{x^2 - y^2} + is$ superconductor, we find $\sigma_{xy}^s = 0$ consistent with
the analysis in the following section on edge states).

The explicit calculation above was restricted to $|\Delta_{xy}| \ll \Delta_0$.
However the result Eq. \ref{shc} holds even away from this limit. This is 
because the system is in the same phase for any finite non-zero value of the 
ratio $\frac{\Delta_{xy}}{\Delta_0}$. The quantized value of the spin Hall 
conductance is a universal property of this phase. A topological invariant 
characterizing the $d+id$
phase has previously been discussed by Volovik\cite{Volovik}. 
The results of this
section provided a physical interpretation of this topological invariance 
in terms of the quantization of the spin Hall conductance. 

\section{Edge states}
\label{edge}
\subsection{Semiclassical argument}
\begin{figure}
\epsfxsize=3.5in
\centerline{\epsffile{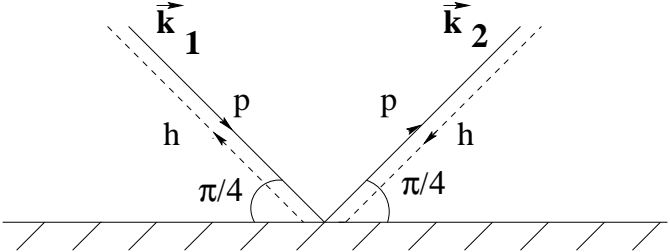}}
\vspace{0.15in}
\caption{Semiclassical trajectory leading to a surface bound state}
\vspace{0.15in}
\label{Andreev}
\end{figure}

As is well known from the theory of the quantum Hall effect, the quantization
of the bulk spin Hall conductance implies the existence, 
for a system with a boundary, of spin-current carrying
states at the edge. In particular $\sigma_{xy}^s = 2$ implies the existence of
two such edge modes.  Consider the $d+id$ superconductor with a boundary, and a
particle incident on the boundary with wave vector $\vec{k}_1$ directed 45 
degrees to the normal.  This particle is reflected to a state with a 
wave vector $\vec{k}_2$ also at 45 degrees to the normal.
This particle can now Andreev reflect off the bulk of the superconductor and
return as a hole (see Fig. \ref{Andreev}). The hole moves on the reverse 
trajectory until it is Andreev reflected from the bulk back as a particle 
at wavevector $\vec{k}_1$.

If the direction of $\vec{k}_1$ corresponds to an angle
$\theta_1$, the direction of $\vec{k}_2$ corresponds to angle $\theta_2 =
\theta_1 \pm \frac{\pi}{2}$.
For the $d+id$ gap $\Delta_k = \Delta_0 \cos(2\theta_k) -i\Delta_{xy}
\sin(2\theta_k)$. Therefore one has
$\Delta_{k_1} = -\Delta_{k_2}$. Thus there is a relative phase shift of $\pi$
for Andreev reflection at $\vec{k}_1$ and $\vec{k}_2$ respectively. The problem
is then formally
identical to that of an SNS junction with a phase shift of $\pi$ between the
two superconductors.
It is well known that in such a system there exists a state at zero energy
bound in the normal layer.
A similar situation obtains if the incident particle is at wave vector 
$-\vec{k}_2$ when again the angle
of incidence is 45 degrees. For all other angles of incidence, the phase shift
for the two Andreev reflections
is different from $\pi$, and there is no bound state. Thus there are precisely
two surface bound states for
every surface orientation of the $d+id$ state. This is entirely consistent with
the quantization of the bulk spin hall conductance to be two.
This is, however, to be contrasted with the $d_{x^2-y^2}$ superconductor
where the existence of such zero energy surface states depends sensitively on
the orientation of the interface\cite{Sauls}. 
Note also that for a $d_{x^2 - y^2} + is$ superconductor,
there is no orientation of the interface for which the phase shift for the two
Andreev reflections is
$\pi$ - hence there are no surface bound states again consistent with the
absence of a quantized spin Hall conductance.

This semiclassical argument can be made precise by solving the
Bogoliubov-de Gennes (B-dG) equations
for the $d +id$ superconductor in the presence of a boundary in the Andreev
approximation.
We remind the reader that the B-dG equations are just the eigenvalue equations
for the $d$-particle
wavefunctions. As the calculations are straightforward, and are very similar to
those in the
literature for the $d_{x^2-y^2}$ superconductor, we will not present them here.
Instead, we will show how the edge modes may be obtained from the
continuum theory described in the previous section.

\subsection{Continuum Dirac theory}
To show the existence of edge states within the effective low energy
Dirac theory, it is necessary that the incident
and reflected modes (at 45 degrees with respect
to the edge) lie along directions in momentum space
which pass close to the nodes of the $d_{x^2-y^2}$ order parameter.
If this is not the case, a description of the edge states requires retaining
bulk modes at high energies of order $\Delta_0$.
To this end, we consider an edge  parallel to the $y$ axis
located at $x=0$.  It is convenient to first rewrite
the Dirac Hamiltonian in the original spatial coordinates ($x,y$):
\begin{eqnarray}
{\cal H} &=& \int d^2x~ \chi^\dagger_1 [-iv\tau^x \partial_x +
-i(v_x \tau^x + v_z \tau^z)\partial_y - \Delta_{xy} \tau^y ]\chi_1  
\nonumber \\
&+& (1 \rightarrow 2, x \rightarrow -x, \Delta_{xy} \rightarrow -\Delta_{xy})~ .
\label{Dirac_xy}
\end{eqnarray}
Here we have performed a rotation about the $\tau^y$
axis by an angle $\theta = \arctan(v_F/v_\Delta)$
and defined $v_x = -v \cos(2\theta)$ and $v_z = v \sin(2\theta)$
with $v^2 = (v_F^2 + v_\Delta^2)/2$.

To establish the appropriate boundary conditions on the $\chi$ fields
at $x=0$, it is necessary to use Eqs. (\ref{d_to_chi1}) and (\ref{d_to_chi2})
re-expressing them in terms of the underlying electron fields.
As emphasized in the previous section, re-expressing
the original BCS Hamiltonian in terms of the
$d$-fermions eliminates all anomalous terms,
reflecting the conservation of spin, even in the presence of the edge.
The appropriate boundary condition is thus simply $d_\alpha(x=0,y) =0$,
which corresponds to the condition:
\begin{equation}
\chi_{1a\alpha}(x=0,y)  = - \chi_{2a\alpha}(x=0,y)  
\end{equation}
on the Dirac fields.

To search for a zero energy edge state it is necessary
to solve the wave equation which follows from the Dirac theory:
\begin{equation}
(-iv\tau^x \partial_x - \Delta_{xy} \tau^y ) \phi(x) = 0  ,
\end{equation}
where we have assumed the (two-component) wave function $\phi_a(x)$
is independent of $y$ - the coordinate {\it along}
the edge.  The appropriate solution which
decays into the sample for $x>0$ is readily found:
$\phi_a(x) = \delta_{a1} \exp(-\Delta_{xy} x/v)$.
At low energies below $\Delta_{xy}$, the Dirac fields can be expanded
in terms of this wavefunction as:
\begin{equation}
\chi_{ja\alpha}(x,y) = (-1)^j~ \sqrt{\frac{\Delta_{xy}}{v}}~ 
\phi_a(x) \chi_{e\alpha}(y) ,
\end{equation}
with a two-component edge Fermion field, $\chi_e(y)$.
Here, the $(-1)^j$ factor has been included to satisfy the
boundary conditions on $\chi(x=0,y)$, and the prefactor under the
square root has been chosen so that the one-dimensional
edge field satisfies canonical anticommutation relations.
The effective edge Hamiltonian can be readily obtained by inserting this 
expansion into the Dirac form in Eq. \ref{Dirac_xy}.  After performing
the $x$-integration one finds:
\begin{equation}
{\cal H}_{edge} = \int dy~ \chi^{\dagger}_{e \alpha} (-iv_e \partial_y )
\chi_{e\alpha} ,
\end{equation}
with edge velocity $v_e = v \sin(2\theta)$.
For the isotropic case this implies $v_e = v_F = v_\Delta$.

The edge Hamiltonian describes a two-component one-dimensional
chiral Fermion.  Each edge mode contributes unity to the dimensionless Hall
conductance, giving $\sigma_{xy}^s = 2$.  Since the edge density operator,
$\chi^\dagger_e \chi_e$, is proportional to the
z-component of spin, this is actually
the {\it spin} Hall conductance, discussed in the previous section.
Rotational invariance of the electron spin 
requires that the Hamiltonian be invariant
under, $\chi \rightarrow i\sigma_y \chi^\dagger$, or equivalently,
\begin{equation}
\chi_e \rightarrow i\sigma_y \chi^\dagger_e .
\end{equation}
The edge Hamiltonian, ${\cal H}_{edge}$, is seen to satisfy this symmetry.
It is instructive to rewrite the edge Hamiltonian back in terms
of the original Dirac fields, $\psi_\alpha$, which
transform as spinors under SU(2) rotations.
In terms of one-dimensional ``edge" Dirac fields defined via,
\begin{equation}
\psi_{e\uparrow} = \chi_{e\uparrow}  ;  \hskip0.4cm
\psi_{e\downarrow} = \chi^{\dagger}_{\downarrow e} ,
\end{equation}
the edge Hamiltonian takes the same form:
\begin{equation}
{\cal H}_{edge}  = \int dy~ \psi^\dagger_e(-iv_e \partial_y ) \psi_{e}   ,
\end{equation}
with an implicit sum on $\alpha$.  
This form is clearly seen to be invariant under
SU(2) rotations:  $\psi_e \rightarrow U \psi_e$,
with $U = \exp(i\bbox{\theta} \cdot \bbox{\sigma})$.

Rather surprisingly, though, the edge Hamiltonian actually
is seen to have an {\it additional} U(1) symmetry;
$\psi_e \rightarrow \exp(i\theta_0) \psi_e$.  This additional symmetry
can be traced to the conserved U(1) ``charge" of the
Dirac $\psi$ particles - called nodons in Ref. \onlinecite{BFN}.  Physically,
this U(1) symmetry reflects the fact that the original BCS Hamiltonian
conserves the {\it difference} between
the number of electrons at one node, say at $\bbox{K}_j$,
and the node with opposite momentum, $-\bbox{K}_j$.  
In the presence of impurities which break momentum conservation, this
additional U(1) symmetry will {\it not}
be preserved.  To see this, consider adding scattering impurities
to the above edge Hamiltonians.  For impurities
which do not break spin rotational invariance,
the edge Hamiltonian must still be invariant under,
$\chi_e \rightarrow i\sigma_y \chi^\dagger_e$,
and, moreover, conserve the z-component of spin:  $\chi^\dagger_e \chi_e$.
A general form satisfying these requirements is,
\begin{equation}
{\cal H}_{imp} = \int dy~ \chi^\dagger_e (\bbox{\eta}(y)
\cdot \bbox{\sigma} ) \chi_e  ,
\end{equation}
where $\bbox{\eta}(y)$ are real functions, random in
the spatial coordinate along the edge.  Rewritten in terms
of the $\psi$ fields these become:
\begin{equation}
{\cal H}_{imp} = \int dy~ [\eta^+ \psi_e \sigma^- \psi_e + H. c. + \eta_z
\psi^\dagger_e \psi_e]  ,
\end{equation}
with $\eta^\pm = \eta_x \pm i\eta_y$ and $\sigma^\pm = (\sigma_x \pm i
\sigma_y)/2$.
Although still invariant under
SU(2) spin rotations $\psi_e \rightarrow U \psi_e$, the additional U(1)
symmetry is clearly {\it not} present anymore,
due to the anomalous ($\psi_e \sigma^- \psi_e$ and $\psi^\dagger_e \sigma^+
\psi^\dagger_e$) terms.

Although the random terms explicitly break the U(1) symmetry,
there is still another {\it hidden} U(1) symmetry, which can be revealed by 
making a clever change of variables.  Specifically, consider defining 
new fields,
\begin{equation}
\tilde{\chi}_e = T_y \exp \left[\frac{i}{v_e} \int^y dy'~ 
\bbox{\eta}(y') \cdot \bbox{\sigma} \right]~ \chi_e,
\end{equation}
where $T_y$ denotes a ``time-ordering" along the spatial coordinate
$y$.  This effectively gauges away the random terms, 
and the full Hamiltonian when
expressed in terms of the new $\tilde{\psi}_e$ fields
exhibits the U(1) symmetry $\tilde{\psi}_e \rightarrow \exp(i\theta_0)
\tilde{\psi}_e$.  This SU(2) gauge transformation will play an important
role in analyzing the network model studied in the next section.

\section{Disorder effects}
\label{phase_dia}
\subsection{Phase diagram}
\begin{figure}
\epsfxsize=3.5in
\centerline{\epsffile{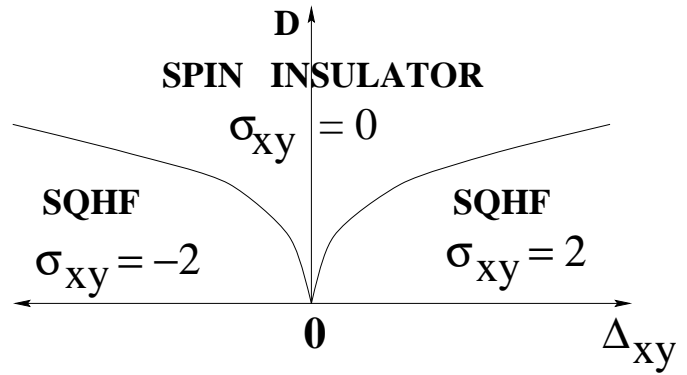}}
\vspace{0.15in}
\caption{Schematic phase diagram as a function of $\Delta_{xy}$ and disorder
$D$; SQHF refers to
the Spin Quantum Hall Fluid}
\vspace{0.15in}
\label{sqh1}
\end{figure}

We now move on to consider the effects of impurities on the $d+id$ 
superconductor.
As shown in the previous section, the edge modes are robust to weak impurity
scattering - hence so is the quantization of the bulk spin Hall conductance. 
Strong impurity scattering can however lead to a transition to a phase with 
zero Hall conductance. It is useful
to consider a phase diagram of the system as a function of $\Delta_{xy}$ and
disorder $D$.
The general topology of such a phase diagram is shown in Fig. \ref{sqh1}.

At zero $D$, $\sigma_{xy}^s = 2~ {\rm sgn}(\Delta_{xy})$. 
This spin quantum hall phase
is stable to weak disorder as seen above. The line $\Delta_{xy} = 0$
is of course the $d_{x^2-y^2}$ superconductor. Turning on disorder at zero
$\Delta_{xy}$
localizes the quasiparticle states at the Fermi energy\cite{short} leading to a
spin insulator.
This phase should be robust to turning on a small $\Delta_{xy}$. This is
particularly clear
in the lattice version of the $d+id$ superconductor in terms of the $d$
operators. The
$\Delta_{xy}$ simply corresponds to a diagonal hopping term, and hence is
innocuous, if weak,
in a localized phase. It is clear then that there must be two transition lines
emerging from the
$D= \Delta_{xy} = 0$ point (symmetrically about the $\Delta_{xy} = 0$ line)
separating the
two quantum Hall phases (with $\sigma_{xy}^s = \pm 2$) from the spin insulator
with $\sigma_{xy}^s = 0$.

Note that the jump in $\sigma_{xy}^s$ is by two\cite{note_2} - 
this is prohibited in generic non-interacting models
of quantum Hall systems but is allowed here due to the special extra $SU(2)$
symmetry.
All phases have zero longitudinal spin conductance. It is interesting to ask
about the behavior of the bulk quasiparticle density of states (DOS)
$\rho(E)$ as a function of energy in various regions of the phase diagram.
It is known\cite{dos} that in the spin insulator without time reversal 
invariance, $\rho(E)$ actually vanishes as $E^2$ at low energies.
In the $d+id$ superconductor, for weak disorder, standard arguments suggest the
development
of exponentially small tails in the density of states leading to a weak filling
in of the gap.
However at disorder strong enough to be near the transition, we expect a larger
density of states
that nevertheless vanishes on approaching zero energy\cite{dos} as $E^2$.

\begin{figure}
\epsfxsize=3.5in
\centerline{\epsffile{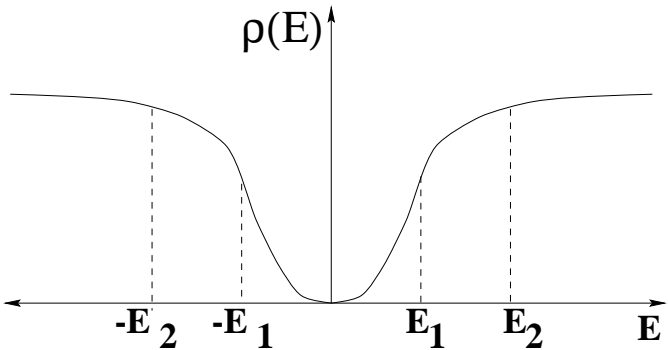}}
\vspace{0.15in}
\caption{Density of states of the $d$ particles showing positions of extended
states}
\vspace{0.15in}
\label{d_ext}
\end{figure}

\begin{figure}
\epsfxsize=3.5in
\centerline{\epsffile{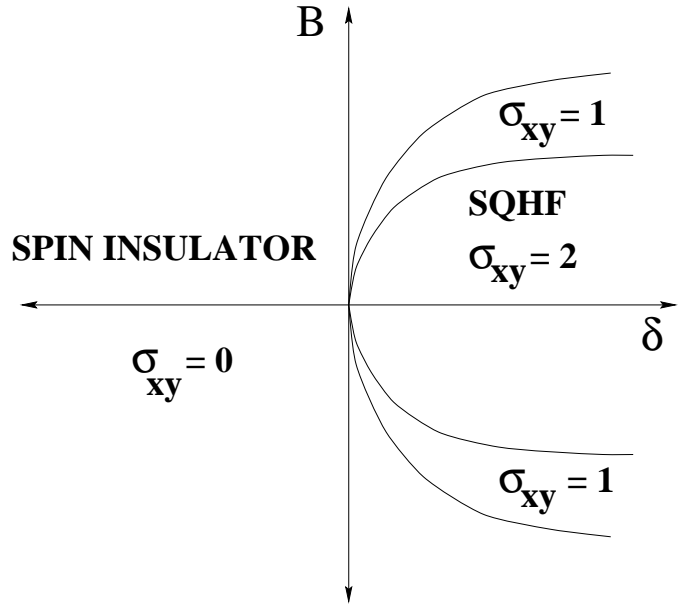}}
\vspace{0.15in}
\caption{Schematic phase diagram as a function of external Zeeman field $B$ (or
energy $E$) and a parameter
$\delta$ measuring the distance (at zero field) from the zero-field phase
boundary for the $0 \rightarrow 2$ transition.}
\vspace{0.15in}
\label{sqh2}
\end{figure}

A different perspective on the phase diagram is provided by considering the
properties of the
wave functions of the single particle Hamiltonian for the $d$ particles.
In the spin quantum Hall phase, $\sigma_{xy}^s = 2$ implies the existence of
precisely two extended
states below the Fermi energy (each contributing unity to $\sigma_{xy}^s$).
These two states will be at two different energies, say $-E_1$ and $-E_2$ (see
Fig. \ref{d_ext}).
The particle-hole symmetry of the $d$ Hamiltonian in Eq. \ref{d_SU2} 
({\it i.e.} the $SU(2)$ spin rotation invariance)
implies the existence of two extended unoccupied states at positive energies
$E_1$ and $E_2$. These states, if filled, contribute $-1$ each to
$\sigma_{xy}^s$.
Thus as we move up in energy and pass $E_1$, $\sigma_{xy}^s$ jumps from $2$ to
$1$ and finally, as we pass
$E_2$, from $1$ to zero. As the disorder increases and we approach the
transition, $E_1$ and $E_2$
collapse towards zero.
A nice way to move up (or down) in energy is by turning on an external Zeeman
field as this acts
exactly like a chemical potential for the $d$ particles. In particular, at
finite Zeeman field,
the transition splits into two separate ones with $\sigma_{xy}^s$ jumping by one
at each. We show in
Fig. \ref{sqh2} the phase diagram in the presence of a Zeeman field.

\subsection{Delocalization Transition}
Let us now consider the properties of the transition (in zero Zeeman field) in
some more detail.
This is a quantum Hall plateau transition where $\sigma_{xy}^s$ jumps by two.
This is a new
universality class for a quantum Hall localization transition distinct from the
usual
one described (for instance) by the Chalker-Coddington network model. A field
theoretic description of this
critical point in two dimensional superconductors without time reversal but with
spin rotation symmetry is obtained on examining the nonlinear sigma model
appropriate
for describing quasiparticle localization in such a system. In a replica
formalism, this is
a sigma model on the space $Sp(2n)/U(n)$\cite{short,dos,Zirn_SC1}. This field
theory admits a
topological term\cite{short} as $\Pi_2(Sp(2n)/U(n)) = Z$ is non-trivial.
We expect by analogy to the reasoning for the conventional integer quantum Hall
transition that the sigma model supplemented with the topological term has a
critical
point which describes the spin quantum Hall transition. Introducing a Zeeman
field induces a crossover to the conventional universality class\cite{short}. 
This is of course consistent
with the transition splitting into two as jumps of $\sigma_{xy}^s$ by more than
one are prohibited in that case.
There is however another very significant difference between the spin quantum
Hall
transition and the conventional one. As mentioned above, the density of states
actually vanishes (at zero energy) on either side of the transition. By
continuity, we expect that the density of states 
vanishes at the critical point as well.

We may now formulate scaling hypotheses for various physical quantities of
interest near the transition. On approaching the critical point (at zero Zeeman
field) by tuning the disorder $D$, for instance, the
localization length $\xi$ (at zero energy) diverges as
\begin{equation}
\xi \sim \delta^{-\nu} .
\end{equation}
where $\delta$ is the distance from the phase boundary.
Moving away from the critical point by turning on a Zeeman field
also introduces a finite localization length $\xi_B$ diverging as
\begin{equation}
\xi_B \sim B^{-\nu_B} .
\end{equation}
We may now obtain, for instance, the behavior of the density of states $\rho(E)$
at the critical point. To that end, note that moving away from zero energy is
the same
perturbation as turning on a Zeeman field. Consequently, the localization
length as a function of
energy diverges as $\xi_E \sim E^{-\nu_B}$. The density of states may now be
obtained by hyperscaling:
\begin{equation}
\rho(E) \sim \frac{1}{E \xi_E^2} \sim E^{2\nu_B - 1} .
\label{hyperscaling}
\end{equation}
For $\delta \neq 0$, $\rho(E)$ satisfies the scaling form
\begin{equation}
\rho(E,\delta) \sim E^{2\nu_B-1}{\bf Y}\left(E
\delta^{-\frac{\nu}{\nu_B}}\right) .
\end{equation}
The universal scaling function ${\bf Y}$ satisfies
\begin{equation}
{\bf Y} (x \rightarrow \infty)  = 1  ,
\end{equation}
\begin{equation}
{\bf Y}(x \rightarrow 0)  \sim  x^{3-2\nu_B} ,
\end{equation}
where the second line follows from requiring that $\rho(E)$ vanishes as $E^2$
off criticality.

\subsection{Network model}
Just as for the conventional quantum Hall transition, it is possible to
construct a network model to describe the universal critical properties.
If we think of the links of the network model as corresponding to
internal edge states of puddles of the quantum Hall fluid immersed in
the spin insulator phase, then it is clear that we need to have two channels of
propagation on each link. The link amplitude is the amplitude of propagation of
the two channels. As the Hamiltonian $H$ describing the dynamics of the system
has the symmetry $\sigma_{y}H^*\sigma_y = -H$, it is clear that the unitary
time evolution operator $U = e^{-iHt}$ satisfies $U^{T}\sigma_y U = \sigma_y$.
Upon restriction to a subspace with $2N$ states, this unitary operator
can be represented by a matrix belonging to the group $Sp(2N)$ (which is
defined precisely
as a $2N \times 2N$ unitary matrix satisfying $U^{T}\sigma_y U = \sigma_y$).
Thus for the case of two channels, the amplitude for propagation is a 
$2 \times 2$ matrix belonging to the group $Sp(2) = SU(2)$. 
The other ingredient in the network model
is the matrix at the node connecting four links.
Formally, this is a scattering event with four incoming channels and
four outgoing channels. The corresponding scattering matrix thus belongs
to the group $Sp(4)$. Taking the link and node scattering
matrices to be random and belonging to $Sp(2)$ and
$Sp(4)$ respectively then completes the specification of the network model.

In some recent work, Kagalovsky et al.\cite{John} have simulated a
network model with these symmetries and obtained numerical estimates
of various critical exponents.
Here however we will follow a different route. We will use the network model to
motivate
the construction of a supersymmetric quantum spin chain which can be used to
calculate various disorder averaged properties of the system. For that purpose,
it is actually more useful to consider an anisotropic version of the network
model in which we view it as a collection of counter-propagating edge modes 
along the $y$-direction.  Two adjacent modes are connected by random tunneling.
(An alternative approach to deriving a superspin chain is discussed in 
Ref. \onlinecite{Ilya}.)
As shown in the previous section, each edge mode is described by a two 
component chiral fermion and is described by the Hamiltonian
\begin{equation}
(-1)^j \int dy~ \chi^{\dagger}_j(y) \left(-i\partial_y + \bbox{\eta}_j(y)
\cdot \bbox{\sigma} \right) \chi_j(y) .
\end{equation}
Here $\chi_j$ refers to the $j$'th edge mode. The $\bbox{\eta}_j(y)$
represent the randomness on the links of the network model. To complete this
Hamiltonian description of the network model, we need to introduce random 
tunneling between neighboring
counter-propagating edge modes. The most general term consistent with the
symmetries required of the Hamiltonian are
\begin{eqnarray}
\sum_j \int dy & \left\{ -it^0_{j}(y) [\chi_{j+1}^{\dagger}(y) \chi_j(y)
- \chi_{j}^{\dagger}(y) \chi_{j+1}(y)] \right.
\nonumber \\ 
& \left. + \vec{t}_j(y) \cdot  [\chi_{j+1}^{\dagger}(y) \vec{\sigma} \chi_j(y) +
\chi_{j}^{\dagger} \vec{\sigma} \chi_{j+1}] \right\}
\end{eqnarray}
Here $t^0_{j}(y)$ and $\vec{t}_{j}(y)$ are random variables with zero mean.

Precisely this Hamiltonian for the case of just two neighboring edge modes
has been studied in detail in Ref. \onlinecite{dos}. It was shown that averages
of physical quantities like the density of states and diffusion propagator
could be obtained from an equivalent supersymmetric quantum mechanical problem
defined by the nonhermitian ``Hamiltonian":
\begin{eqnarray}
\label{superH}
h & = & h_{ff} + h_{bb} + h_{fb} + h_{\omega} \\
h_{ff} & = & -J \left[(f_1^{\dagger} \sigma_y f_1^{\dagger})
(f_2^{\dagger} \sigma_y f_2^{\dagger}) \right.
\nonumber \\
&    & + \left. (f_1 \sigma_y f_1)(f_2 \sigma_y f_2) 
+ 2(f_1^{\dagger} f_1 - 1)(f_2^{\dagger}f_2 - 1)\right]
 \\
h_{bb} & = & 2J(b_1^{\dagger} b_1 + 1) (b_2^{\dagger} b_2 + 1) \\
h_{fb} & = & 2J \left[(b_1^{\dagger} \sigma_y f_1^{\dagger})
(f_2^{\dagger} \sigma_y b_2^{\dagger}) + (b_1 \sigma_y f_1) (f_2\sigma_y b_2)
\right. \nonumber \\
&  & - \left. (f_1^{\dagger} b_1) (f_2^{\dagger} b_2) -
(b_1^{\dagger} f_1) (b_2^{\dagger} f_2)) \right]    \\
h_{\omega} & = & \omega (f^{\dagger}_1 f_1 + b^{\dagger}_1 b_1
+ f^{\dagger}_2 f_2 + b^{\dagger}_2 b_2)
\end{eqnarray}
Here $f_j$ $(b_j)$ are two component fermionic (bosonic) operators, and
index $j = 1, 2$ labels the two edge modes.  Parameter
$\omega$ is the imaginary frequency at
which we wish to compute averages. The constant $J > 0$ is determined by
the strength of the disorder. Its actual value is unimportant for calculation
of universal properties. We refer the reader to Ref. \onlinecite{dos} for 
further details.  In the following we set $J = 1$ for convenience. 
This superHamiltonian generates time evolution in the $y$-direction.
Clearly the superHamiltonian describing the full network can be built up from
this two edge Hamiltonian.  Just as in the case of 
the superspin chain which describes the conventional quantum Hall 
transition\cite{Zirn},
the two distinct phases on either side of the transition correspond to the two
possible ways of dimerizing
the chain. The critical point corresponds to the uniform chain where the bond
strength $J$ is the same for all bonds.  An important feature of this
Hamiltonian that is not shared by superspin chains constructed for the
conventional quantum Hall transition\cite{Zirn,Tsai} 
is that the low energy sector of this theory is described by a finite 
on-site Hilbert space\cite{dos} of dimension $D = 3$:  
\begin{eqnarray}
| 1 \rangle &\equiv& | 0 \rangle
\nonumber \\  
| 2 \rangle &\equiv& \frac{1}{\sqrt{2}}~ \epsilon_{\alpha \beta} 
b^{\dagger}_{\alpha} f^{\dagger}_{\beta} | 0 \rangle
\nonumber \\
| 3 \rangle &\equiv& \frac{1}{2}~ \epsilon_{\alpha \beta} 
f^{\dagger}_{\alpha} f^{\dagger}_{\beta} | 0 \rangle~ .
\end{eqnarray}
This crucial simplification permits considerable numerical 
and analytical progress.

For a chain of $L$ (even) sites, the superHamiltonian may be 
written, following the notation of Ref. \onlinecite{Tsai}, as:
\begin{eqnarray}
H &=& \sum_{j=0}^{L-2}~ J_j~
\left[ \sum_{a=1}^4~ g_a~ S^a_j~ S^a_{j+1}
 + (-1)^j~ \sum_{a=5}^{8}~ g_a~ S^a_j~ S^a_{j+1} \right]
\nonumber \\
&& + \sum_{j=0}^{L-1} \omega_j~ \left[ S^1_j + S^2_j \right]\ .
\label{H-SUSY}
\end{eqnarray}
Here $J_j = [1+(-1)^j \delta]$ where the relevant dimerization parameter 
$\delta = 0$ 
at the critical point.  We have introduced different imaginary frequencies 
$\omega_j$ at each site to permit the extraction of critical properties (see
the following section).  The constants $g_a$ are defined to be:
\begin{eqnarray}
g_a = \left\{ \begin{array}{ll} 2;~ a = 1, 7, 8\\
\\
1;~ a = 3, 4 \\ 
\\
-2;~ a = 2, 5, 6 \ .
\end{array}\right.
\label{signs}
\end{eqnarray}
In Eq. \ref{H-SUSY} we have introduced 8 spin operators:
\begin{eqnarray}
\begin{array}{l}
S^1 \equiv b^{\dagger}_{\alpha} b_\alpha + 1 \\ \\
S^2 \equiv f^{\dagger}_{\alpha} f_\alpha - 1 \\ \\
S^3 \equiv \epsilon_{\alpha \beta} f^{\dagger}_{\alpha} f^{\dagger}_{\beta} \\ \\
S^4 \equiv \epsilon_{\alpha \beta} f_\alpha f_\beta \\ \\
\end{array} \ \ \
\begin{array}{l}
S^5 \equiv \epsilon_{\alpha \beta} b^{\dagger}_{\alpha} f^{\dagger}_{\beta} \\ \\
S^6 \equiv \epsilon_{\alpha \beta} b_\alpha f_\beta \\ \\
S^7 \equiv b^{\dagger}_{\alpha} f_\alpha \\ \\
S^8 \equiv f^{\dagger}_{\alpha} b_\alpha \\ \\
\end{array}
\label{superspins}
\end{eqnarray}
Bosonic-valued operators $S^1, \ldots, S^4$ make up the symmetric sector of the
Hamiltonian while fermion-valued operators $S^5, \ldots, S^{8}$ are in the
antisymmetric sector.  Despite the fact that $H$ is non-Hermitian, it
only has real-valued eigenvalues.  $H$ is also defective (the left eigenstates
or the right eigenstates do not separately span the whole Hilbert space), 
complicating the numerical problem of diagonalizing it. 

The Hamiltonian commutes with two (fermion-valued) supersymmetry generators,
$[H,~ Q_1] = [H,~ Q_2] = 0$, where
\begin{eqnarray}
Q_1 &\equiv& \sum_j \bigg{[} b^{\dagger}_{j\alpha} f_{j \alpha}
- (-1)^j f^{\dagger}_{j\alpha} b_{j \alpha} \bigg{]} .
\nonumber \\
Q_2 &\equiv& \sum_j \bigg{[} (-1)^j b^{\dagger}_{j\alpha}
f_{j \alpha} + f^{\dagger}_{j\alpha} b_{j \alpha} \bigg{]} .
\label{SUSY-charges}
\end{eqnarray}
It is not difficult to see that the supersymmetric Hamiltonian must have a
unique, zero-energy, ground state.  The right and left (ground) eigenstates
are therefore annihilated by the Hamiltonian:
$H | \Psi_0 \rangle = \langle \Psi_0 | H = 0$.  Also, the
ground state is annihilated by the SUSY charges:
$Q_1 | \Psi_0 \rangle = Q_2 | \Psi_0 \rangle = 0$.
All excited states appear in quartets or
larger multiples of 4, half with odd total fermion content,
and these cancel out in the partition function by virtue
of the supertrace:
\begin{equation}
Z = {\rm STr} e^{-\beta H} \equiv {\rm Tr} (-1)^{N_f} e^{-\beta H} = 1,
\label{partition}
\end{equation}
where $N_f$ is the total number of fermions.

\section{DMRG results}

We employ the relatively simple ``infinite-size'' DMRG algorithm\cite{White} to 
numerically access the properties of the critical point $\delta = 0$. 
The fact that the ground state energy is exactly zero provides a valuable check 
on the accuracy of the DMRG algorithm which incurs errors when, as the chain 
length increases, the Hilbert spaces of the blocks grow beyond the finite 
limit of $M$ states.  Increasing $M$ up to limits set by machine memory and 
speed yields systematic improvement in the accuracy of the DMRG
algorithm.  In results reported below we have checked that $M$ is sufficiently
large to ensure adequate accuracy; for $M \geq 243$ there is no
truncation until the chain exceeds length $L = 12$.  Reasonable accuracy is 
maintained, for the case $M = 256$, out to $L = 26$: 
the ground state, when targeted, has an energy $E_0$ which increases from zero 
to just $E_0 = 2.3 \times 10^{-4}$ at $L = 26$.  Furthermore, for $M = 512$, 
the ground state energy is only $E_0 = 3.2 \times 10^{-5}$ 
at chain length $L = 30$, showing the systematic improvement in accuracy with
increasing $M$.

Reduced density matrices for the two augmented blocks,
each of Hilbert space size $D \times M$, are formed by
computing a partial trace over half the chain.
For the left half of the chain the density matrix is chosen to have
the following symmetric form\cite{Jane}:
\begin{equation}
\rho_{i j} = \frac{1}{2}~ \sum_{i^\prime=1}^{D M}~ \left\{
\Psi^L_{i i^\prime}~ \Psi^L_{j i^\prime} +
\Psi^R_{i i^\prime}~ \Psi^R_{j i^\prime} \right\}~ ;
\label{density}
\end{equation}
a similar formula holds for the right half of the chain.
Here $\Psi^R_{i i^\prime} \equiv \langle i, i^\prime | \Psi \rangle$
and $\Psi^L_{i i^\prime} \equiv \langle \Psi | i, i^\prime \rangle$
are, respectively, the real-valued matrix elements of the targeted
right and left eigenstates projected onto a basis
of states labeled by unprimed Roman index $i$ which covers the left half
of the chain and primed index $i^\prime$ which covers the right half.
To compute ground state properties, $\Psi$ is selected to be the ground state;
conversely, to find the gap, $\Psi$ is chosen to be one of the lowest-lying
excited states.  All of the eigenvalues of $\rho$ are real and positive;
these are interpreted as probabilities and the $(D-1) M$ least probable
states are thrown away.

\begin{figure}
\epsfxsize=3.5in
\centerline{\epsffile{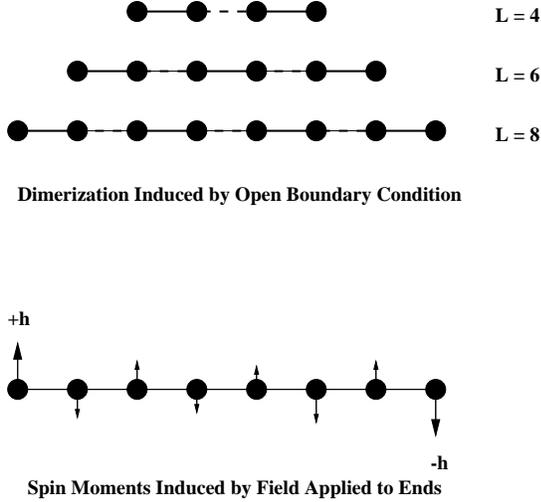}}
\vspace{0.15in}
\caption{Extraction of critical behavior from finite-size effects, illustrated
for the case of an ordinary quantum antiferromagnetic spin chain.  
Dimerization of the nearest-neighbor spin-spin correlation function, indicated
here by alternating strong (solid) and weak (dashed) bonds, 
is induced by the open boundary conditions.  Spin moments are 
induced by the application of a magnetic field of strength $\pm h$ to the 
two spins at the ends of the chain (shown) or formed by the 
application of a staggered field throughout the chain.}
\vspace{0.15in}
\label{dimer}
\end{figure}

To extract critical behavior, we monitor the induced 
dimerization and spin moments near the
center of the chain as the chain length $L$ is enlarged via the DMRG 
algorithm\cite{Jane}.  Dimerization is induced by the open boundary conditions
as shown in Fig. \ref{dimer}. 
Spin moments are formed in the interior of the chain in two different ways.  
In the bulk case $\omega_j$ is set equal to a small, but non-zero, constant
$\omega > 0$ on each site, inducing non-zero spin moments.  Alternatively, the 
spins at the chain ends can be fixed by setting $\omega_j = 0$ except at the
chain ends where $\omega_j$ is assigned a large value which completely 
polarizes the end spins, see Fig. \ref{dimer}.  Power-law scaling of the 
induced dimerization and spin moments in the interior of the chain is 
expected\cite{scaling} at the critical point $\delta = 0$.  
As discussed earlier, we may move off criticality either by dimerizing the spin
chain or by turning on a finite Zeeman field 
(which is equivalent to going away from zero energy).
There are two independent exponents related to these two perturbations of the 
critical spin chain. As in Section \ref{phase_dia}, we may write down 
scaling forms for various physical quantities. For a finite system size, these 
scaling forms will involve
two scaling variables: the ratio $\frac{\xi}{\xi_B}$ of the two 
localization lengths and the ratio
$\frac{\xi}{L}$.  Consider, for instance, the density of states. 
This is determined by the boson occupancy according to\cite{dos}
\begin{equation}
2 \pi \rho(E)  =  \frac{2}{L}Re \sum_i \langle 1 + b^{\dagger}_i b_i \rangle
\end{equation}
where we calculate expectation values setting 
$\omega_j = \omega =-i( E + i\eta)$.
Thus $\rho(E)$ can be obtained from the behavior of the spin operator $S^1$.
This scales at the center of the
chain as a function of the chain length $L$ and the uniform, ``bulk,'' 
imaginary frequency $\omega_j = \omega$ as follows:
\begin{equation}
\langle S^1_{L/2} \rangle = \omega^\alpha~ 
f(L \omega^{\nu_B}, \xi \omega^{\nu_B})
\end{equation}
where the exponent    
\begin{equation}
\alpha = 2 \nu_B - 1\ ,
\end{equation}
as required by  hyperscaling (see Eq. \ref{hyperscaling}). 
When the applied dimerization $\delta = 0$, this reduces to 
\begin{eqnarray}
\langle S^1_{L/2} \rangle &=& \omega^\alpha~ g(L \omega^{\nu_B})
\nonumber \\
&\sim& L^y \omega\ \ \ \ \ {\rm as}~ \omega \rightarrow 0~ .
\label{occ}
\end{eqnarray}
Here the scaling function $g(x)$ is given, for $|x| \ll 1$, by:
\begin{equation}
g(x) = x^{-\alpha/\nu_B}~ (c_1 x^{1/\nu_B} + c_2 x^{2/\nu_B} + \ldots)\ .
\end{equation}
This equation expresses the fact that when the system length is much smaller
than the correlation length ($|x| \ll 1$),
the DOS is an analytic, linear, function of the imaginary
energy $\omega$.  With this scaling form we obtain
\begin{equation}
y =2 (1-\nu_B)/\nu_B\ .
\end{equation}

In what follows, we first describe the calculation of the 
exponents $\nu, \nu_B$ for the two diverging localization
lengths. These can then be used to extract the other critical 
exponents $\alpha,~ y$ using the above scaling arguments. 
We will however provide independent support for the validity of 
these scaling arguments by direct calculation.  

1. Localization length exponent $\nu_B$.  The localization length scales,
as a function of the imaginary frequency $\omega$, with exponent $\nu_B$:
\begin{equation}
\xi_\omega \sim \omega^{-\nu_B}\ .
\end{equation}
One way to determine $\nu_B$ is to find the crossover, for uniform 
$\omega_j = \omega > 0$,
from power law decay of the induced dimerization to exponential decay.
The induced dimerization at the center of the chain is defined as
\begin{equation}
\Delta(L, \omega) \equiv | \langle S^3_{L/2-1} S^3_{L/2}
- S^3_{L/2} S^3_{L/2+1} \rangle |~ ,
\end{equation}
where we recall that
$S^3 \equiv \epsilon_{\alpha \beta} f^{\dagger}_{\alpha} f^{\dagger}_{\beta}$
is one of the 8 SUSY spin operators (each of the 7 other spin operators scale
similarly.)  It has the following asymptotic behavior:
\begin{eqnarray}
\Delta(L, \omega) = \left\{ \begin{array}{ll} C~ L^{-x};\ \ \ \omega = 0\\
\\
C^\prime~ e^{- m(\omega) L}; \ \ \ m(\omega) L \gg 1 .
\end{array}\right.
\label{dimer-scale}
\end{eqnarray}
Fits to the second line in Eq. \ref{dimer-scale}
permit the extraction of the mass gap $m(\omega)$; then $\nu_B$ is determined 
by a power law fit to $m(\omega) \sim \omega^{\nu_B}$.
We find $\nu_B = 0.55 \pm 0.1$ for calculations with $M = 128$, fitting
over the range $20 \leq L \leq 24$ and $0.1 < \omega < 0.8$.
A direct calculation of the gap in the excitation spectrum as a function of
$\omega$ also yields results consistent with this value for $\nu_B$. Note 
that the exact result of Ref. \onlinecite{Ilya} is 
$\nu_B = 4/7 = 0.5714... $. 

\begin{figure}
\epsfxsize=3.5in
\centerline{\epsffile{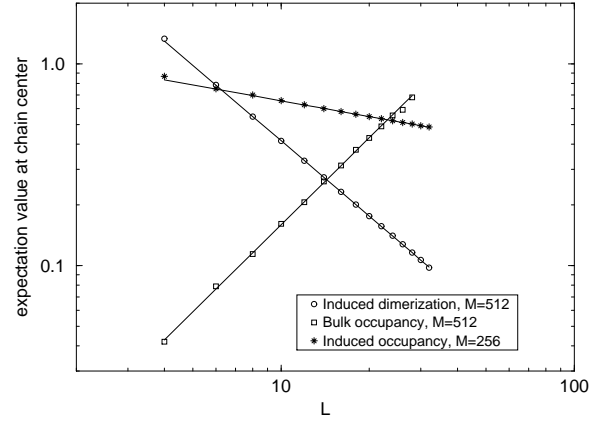}}
\vspace{0.15in}
\caption{Power-law scaling of the induced dimerization, the bulk occupancy, and 
the induced occupancy with chain length $L$.  
In the case of the induced dimerization and
the bulk occupancy, $\omega_j = 10^{-5}$ throughout the chain, small 
enough for the bulk occupancy 
to be well described by the second line of Eq. \ref{occ}.  The bulk
occupancies have been multiplied by a factor of $10^3$. 
The induced occupancy is obtained by setting
$\omega_j = 0$ everywhere except at the chain ends where it is made large, in
this case $\omega_0 = \omega_{L-1} = 10$.  Straight lines are fit to each of 
the three data sets.}
\vspace{0.15in}
\label{scale}
\end{figure}

2. Dimerization exponent $\nu$.  For small $\omega_j = 10^{-5}$,
fitting the induced dimerization shown in Fig. \ref{scale} to the first line 
in Eq. \ref{dimer-scale} yields $x = 1.24 \pm 0.01$.  The dimerization 
exponent $\nu$ is related to the scaling dimension $x$ by 
\begin{equation}
\nu = \frac{1}{2 - x}
\end{equation}
and thus $\nu = 1.32 \pm 0.02$, 
close to the percolation value of $4/3$ reported in Ref. \onlinecite{Ilya}.  
For the 
spin-$1/2$ Heisenberg antiferromagnet and the spin-1 antiferromagnet at the
critical point accuracy at the few percent level was also achieved\cite{Jane}.
The network model simulations\cite{John} find $\nu \simeq 1.12$. Though this 
is close to the value we find numerically, and to the exact 
result\onlinecite{Ilya}, the reason for the
lack of more precise agreement is unclear to us.

3. DOS exponents $\alpha$ and $y$.   
Drawing upon the data shown in Fig. \ref{scale} we obtain
$y = 1.43 \pm 0.05$ by direct fit of the bulk occupancy at one of 
the central sites to the second line of Eq. \ref{occ}.  
The error is estimated by comparing 
results from DMRG calculations with $M = 256$ and $M = 512$ and also by making 
power-law fits over different ranges of chain lengths $L$. 
This calculation of $y$ can now be used to calculate $\nu_B = 0.58 \pm 0.01$ 
in good agreement with the value obtained in item 1 above. 

As mentioned above, the scaling of the DOS can be extracted in another way: 
set $\omega_j = 0$ everywhere along the chain except at the two sites at the 
ends of the chain where it is made large. 
Consequently at the chain ends $\langle S^1_0 \rangle = 
\langle S^1_{L-1} \rangle = 1$ but in the interior
the expectation value $\langle S^1_{L/2} \rangle$, which we call 
the induced occupancy, decreases as the chain grows in length:
\begin{equation}
\langle S^1_{L/2} \rangle \sim \frac{1}{L^w}\ .
\end{equation}
From Fig. \ref{scale} we find $w = 0.26 \pm 0.02$.
Now, scaling relates $L \sim \omega^{-\nu_B}$ and hence
$\alpha = w \nu_B$. Using the relation $\alpha = 2 \nu_B - 1$, we get
$\nu_B = 0.57 \pm 0.02$ again in agreement with the estimates above, 
and the exact result\onlinecite{Ilya}. 
Note that the density of states exponent $\alpha = 0.14 \pm 0.04$.  

\section{Discussion}
How may the physics discussed in this paper be probed if a 
$d_{x^2 - y^2} + id_{xy}$ superconductor were to be found experimentally? 
The bulk of this paper has focused
on spin Hall transport which is extremely difficult to measure. However, the
thermal Hall
conductance is also quantized in the $d_{x^2 - y^2} + id_{xy}$ state. This can,
for instance, be seen
using the edge state theory developed in Section \ref{edge}. Indeed, if the
temperature of
one edge is raised by $\delta T$ relative to the other, the excess heat current
is easily seen
to be $\frac{2\pi^2 T \delta T k_B^2}{3 h}$ implying a thermal Hall conductance
of
\begin{equation}
\kappa_{xy} = \frac{2\pi^2 T k_B^2}{3h}
\end{equation}
Thus $\frac{\kappa_{xy}}{T}$ is quantized\cite{KF} in the $d_{x^2 - y^2} + 
id_{xy}$ superconductor.
On the other hand, in the spin insulator phase, $\frac{\kappa_{xy}}{T}$
goes to zero as the temperature goes to zero.
Note that the charge Hall conductance is {\em not} quantized in the $d+id$
phase\cite{GI}.
Physically this is because any edge quasiparticle electrical current causes
flow of supercurrent in the opposite direction out to a distance of order the
penetration depth.

The behavior of the quasiparticle density of states may be probed by specific
heat, spin susceptibility,
or tunneling measurements. We caution, however, that it may be necessary to
include quasiparticle
interactions, neglected in the theory so far, to obtain meaningful comparisons
with experiments
for these quantities. (The quantization of the spin and thermal Hall
conductances is expected
to be robust to inclusion of quasiparticle interactions).

It is interesting to ask about experimental realizations of $d+id$ pairing
symmetry in
layered three dimensional superconductors. If each layer is deep in the spin
quantum Hall fluid
phase, then arguments similar to those for multilayer quantum Hall
systems\cite{cmet_t,cmet_e}, imply the existence of a
``chiral spin metal" phase at the surface with diffusive spin transport in the
direction perpendicular
to the layers and ballistic spin transport within each layer. The properties of
this chiral spin metal
will be quite similar to those of the chiral metal discussed in multilayer
quantum Hall systems\cite{cmet_t,cmet_e}.

Throughout this paper, we have analyzed only the case of spin singlet pairing.
For triplet pairing, such as in a $p$-wave superconductor,
neither the spin nor the charge of the quasiparticles is conserved. Thermal
transport still remains a useful way of probing quasiparticle transport. 
Arguments very similar to those
used in this paper show that a two dimensional superconductor with $p_x +i p_y$
symmetry has a quantized thermal Hall conductance. For a layered three 
dimensional system,
we then have a chiral surface sheath with diffusive thermal transport in the
direction
perpendicular to the layers, and ballistic thermal transport within each layer.
Such a layered $p+ip$ superconductor is possibly realized in the material
$Sr_2RuO_4$\cite{ruth}.

We thank Leon Balents, John Chalker, Steve Girvin, Ilya Gruzberg, Andreas
Ludwig, Chetan Nayak, Nick Read, Shan-Wen Tsai, and Xiao-Gang Wen  
for useful discussions.  This research was supported by NSF Grants DMR-9704005,
DMR-9528578, DMR-9357613, DMR-9712391, and PHY94-07194.
Computations were carried out with double-precision C++ code on Cray PVP
machines at the Theoretical Physics Computing Facility at Brown University.

\end{multicols}

\begin{references}
\bibitem{note_1} We consider only the situation in which the gap is much
smaller than the Fermi energy.
There is actually another phase (``molecular limit") in which the gap is so
large that, at fixed density,
the chemical potential has actually gone negative. Such a phase does not have a
quantized spin Hall conductance.
\bibitem{note_2} This was also emphasized in Ref. \onlinecite{John} using the
network model.
\end{references}
\end{document}